\documentclass{iaus}
\usepackage{graphicx}

\title[Galactic Plane Survey] %% give here short title %%
{Missing Galactic PNe: [S\,{\large III}] Imaging Survey}

\author[J. Shiode et al.]   %% give here short author list %%
{J. Shiode, D. P. Clemens, K. A. Janes \and A. Pinnick}

\affiliation{Institute for Astrophysical Research, Boston Univ., 725
Comm. Ave., Boston, MA, USA}

\pubyear{2006}
\volume{234}  %% insert here IAU Symposium No.
\pagerange{1--2}
\date{?? and in revised form ??}
\jname{Planetary Nebulae in our Galaxy and Beyond}
\editors{M. J. Barlow \& R. H. M\'endez, eds.}
\begin{document}

\maketitle

\begin{abstract}
The total number of Galactic planetary nebulae (GPNe) is highly uncertain; 
the most inclusive current catalog contains only $\sim$ 1,500. We will use the PRISM wide-field 
imager on the 1.83 m Perkins Telescope to conduct a pilot survey 
of the Galactic plane in search of \mbox{[S \textsc{iii}]} emission from PNe obscured by dust and 
missed by surveys of H$\,\alpha$. We are employing the method of 
Jacoby \& Van de Steene (JVS), who surveyed the bulge for \mbox{[S \textsc{iii}]} $\lambda$9532.
In addition to seeing through more of the extinction, use of the 
\mbox{[S \textsc{iii}]} emission line will \textit{a priori} reject the most troublesome catalog 
contaminants: ultracompact \mbox{H$\;$\textsc{ii}} regions.
\keywords{Surveys, Catalogs, Galaxy: disk}
\end{abstract}

\begin{figure}[h]
    \centering
    \includegraphics[width=0.5\textwidth]{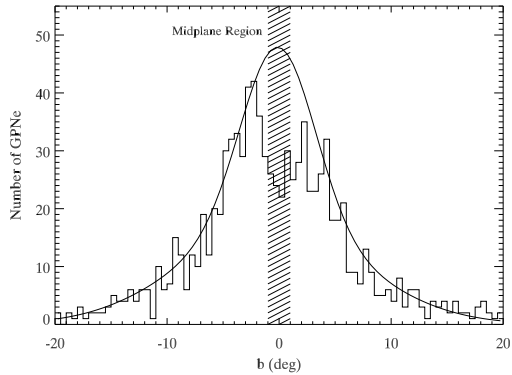}
    \caption{Numbers of known disk PNe binned by Galactic
    latitude.} \label{fig:hist}
\end{figure}

Current estimates of the number of Galactic Planetary Nebulae (GPNe) in the
Milky Way range from a few thousand (\cite[de Marco \& Moe 2005]{deMarco05})
to several tens of thousands (\cite[Frew 2006]{Frew06}). The actual number of
observed and verified GPNe is just $\sim$ 3,000 (\cite[Parker \etal\ 2003]{Parker03}).
Figure~\ref{fig:hist} is an updated version of the histogram
compiled by \cite[Kistiakowsky \& Helfand (1993)]{Kistiakowsky93}, showing the Galactic
latitude distribution of GPNe in the Catalogue of Galactic
Planetary Nebulae (\cite[Kohoutek 2001]{Kohoutek01}). The sample includes all GPNe at low
Galactic latitude, and $\left|\ell\right|>10\,^{\circ}$ (to exclude
the Galactic bulge). There is a clear deficit in the number of known GPNe near $b=0\,^{\circ}$, 
most likely due to obscuration by dark clouds. This low-latitude 
region is not easy to survey for GPNe using the standard \mbox{[O \textsc{iii}]} or 
H$\alpha$ signposts because of the large extinctions.

The \mbox{[S \textsc{iii}]} $\lambda$9532 emission line is seen in most PNe. Ultra-compact
\mbox{H$\;$\textsc{ii}} regions (\mbox{UCH$\:$\textsc{ii}} regions) are the most numerous 
contaminants of current GPNe catalogs. Detectable \mbox{[S \textsc{iii}]} $\lambda$9532 
emission is expected from only the hottest and most massive \mbox{UCH$\;$\textsc{ii}} regions 
due to the high ionization levels needed to produce and sustain \mbox{S$\,^{++}$}. 
The \mbox{[S \textsc{iii}]} $\lambda$9532 line has been used at least twice before to look
for PNe. \cite[Kistiakowsky \& Helfand (1993)]{Kistiakowsky93}
looked for \mbox{[S \textsc{iii}]} $\lambda$9532 emission from compact 20 cm radio sources
in the Galactic plane, finding \mbox{[S$\;$\textsc{iii}]} emission toward
10 of 11 candidate PNe, but none associated with candidate SNRs. \cite[JVS]{Jacoby04} 
surveyed the center \mbox{$4\,^{\circ}\times\,4\,^{\circ}$} of the Milky Way for extincted PNe, finding
94 new candidates---roughly 6 per square degree. Each group found
\mbox{[S \textsc{iii}]} $\lambda$9532 to be the most prominent line for V-band extinctions of
4 to 12 mag.

\begin{figure}[h]
    \centering
    \includegraphics[width=0.8\textwidth]{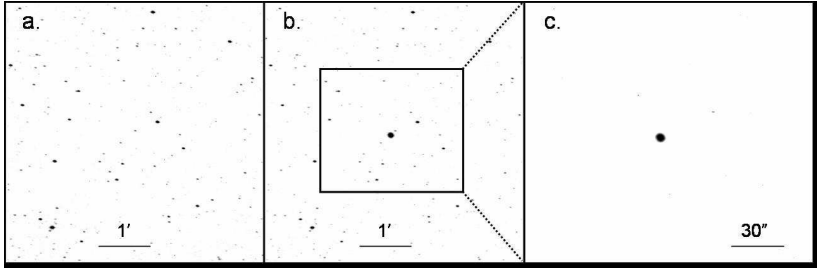}
  \caption{Test field in the Galactic plane, centered on PNG326.7+00.7. 
   \textbf{a.)} off-band image \textbf{b.)} on-band image \textbf{c.)} an enlarged portion of the 
   on-minus-off differenced image.}\label{fig:image}
\end{figure}

Our survey uses two 20$\;$\AA\ narrow bandpass interference
filters to sample the emission line and spectrally adjacent continuum. The
on-band filter rejects the Paschen 8 recombination line at 9546$\;$\AA, while avoiding 
telluric OH lines. The off-band filter is close in wavelength while also avoiding 
telluric OH. Two matched filters, placed close together, produce 
differenced images able to accurately identify \mbox{[S$\;$\textsc{iii}]}
emission sources, and thus candidate GPNe (see Fig.~\ref{fig:image})

Observations will be conducted during Summer (and possibly Fall) 2006 using
the new \textit{Perkins Re-Imaging System (PRISM)} on the Perkins 1.83 m
telescope in Flagstaff, AZ. Figures~\ref{fig:image}a and~\ref{fig:image}b
display PRISM's $10\,'\times\,10\,'$ effective field-of-view. Five minute
integrations through each filter for each field allow us to reach
the same depth as the \cite[JVS]{Jacoby04} survey. We will observe 
the Galactic plane in one-field-wide latitude strips
from $b=-1\,^{\circ}$ to $b=+1\,^{\circ}$. Some known GPNe and 
\mbox{H$\;$\textsc{ii}} regions will be sampled for calibration. 

The images obtained through each filter will be calibrated and
on-minus-off differenced images will be computed. We will visually inspect
the differenced images in search of resolved \mbox{[S \textsc{iii}]} emission,
compile point source catalogs with pairs of filter magnitudes, and
perform magnitude differencing in search of unresolved
\mbox{[S$\;$\textsc{iii}]} $\lambda$9532 emission. The resulting list of candidate 
GPNe will be cross-referenced with current H$\alpha$ and centimeter-wavelength radio 
surveys for confirmatory evidence about the candidate sources. We expect that
verified GPNe selected as candidates using the \mbox{[S \textsc{iii}]} $\lambda$9532
emission line will form a deep, minimally contaminated list of Galactic
Planetary Nebulae.
\newline

\textit{We thank George Jacoby for his help and advice.
JS is supported by the NSF-REU Program under grant AST--0440936--1.
PRISM was developed with support from the NSF (AST--0079541, KAJ, P. I.), Boston Univ. and Lowell Obs.}

\end{document}